\documentstyle[psfig,natbib209]{mn}

\def\ifundefined#1{\expandafter\ifx\csname#1\endcsname\relax}

\def\la{\mathrel{\hbox{\rlap{\hbox{\lower4pt\hbox{$\sim$}}}\hbox{$<$}}}}
\def\ga{\mathrel{\hbox{\rlap{\hbox{\lower4pt\hbox{$\sim$}}}\hbox{$>$}}}}

\newcommand{\be}{\begin{eqnarray}}
\newcommand{\ee}{\end{eqnarray}}

\ifundefined{nuc}\def\nuc#1#2{\relax\ifmmode{}^{#1}{\protect\text{#2}}
\else${}^{#1}$#2\fi}\else\relax\fi

\newcommand{\etal}{et al.}

\newcommand{\msol}{\ifmmode{{\rm M}_\odot}\else{M$_\odot$}\fi}
\newcommand{\foe}{\ifmmode{10^{51}}\else{$10^{51}$}\fi}
\newcommand{\nni}{\nuc{56}{Ni}}
\newcommand{\xni}{\ifmmode{{\rm X}_{\rm Ni}}\else{X$_{\rm Ni}$}\fi}
\def\ang{\hbox{\AA}}
\def\Teff{\ifmmode{T_{\rm eff}}\else{\hbox{$T_{\rm eff}$} }\fi}
\def\Tmod{\ifmmode{T_{\rm model}}\else{\hbox{$T_{\rm model}$} }\fi}
\def\Rzero{\ifmmode{R_0}\else{\hbox{$R_0$} }\fi}
\newcommand{\vno}{\ifmmode{v_0}\else{\hbox{$v_0$} }\fi}

  at 12.0 true pt 

\def\etal{{et al}}

\def\g{\gamma}
\def\b{\beta}
\def\m{\mu}

\def\n{\nu}

\def\pder#1#2{{\partial #1 \over \partial #2}}
\def\div#1#2{{#1\over #2}}
\def\rout{\ifmmode{r_{\rm out}}\else\hbox{$r_{\rm out}$}\fi}
\def\tmax{\ifmmode{\tau_{\rm max}}\else\hbox{$\tau_{\rm max}$}\fi}
\def\tstd{\ifmmode{\tau_{\rm std}}\else\hbox{$\tau_{\rm std}$}\fi}
\def\vmax{\ifmmode{v_{\rm max}}\else\hbox{$v_{\rm max}$}\fi}
\def\muE{\ifmmode{\mu_{\rm E}}\else\hbox{$\mu_{\rm E}$}\fi} 
\def\pE{\ifmmode{p_{\rm E}}\else\hbox{$p_{\rm E}$}\fi} 
\def\bmax{\ifmmode{\b_{\rm max}}\else\hbox{$\b_{\rm max}$}\fi}
\def\kms{\hbox{$\,$km$\,$s$^{-1}$}}

\def\ang{\hbox{\AA}}

\def\Teff{\hbox{$\,T_{\rm eff}$} }

\def\rout{\hbox{$r_{\rm out}$} }

\def\chistd{\ifmmode{\chi_{\rm std}}\else\hbox{$\chi_{\rm std}$}\fi}

\def\lstar{\ifmmode{\Lambda^*}\else\hbox{$\Lambda^*$}\fi} 
\def\Rop{\ifmmode{[R_{ij}]}\else\hbox{$[R_{ij}]$}\fi}

\def\Rji{\ifmmode{[R_{ji}]}\else\hbox{$[R_{ji}]$}\fi}
\def\Rstar{\ifmmode{[R_{ij}^*]}\else\hbox{$[R_{ij}^*]$}\fi}

\def\Rjistar{\ifmmode{[R_{ji}^*]}\else\hbox{$[R_{ji}^*]$}\fi}
\def\DRji{\ifmmode{[\Delta R_{ji}]}\else\hbox{$[\Delta R_{ji}]$}\fi}
\def\DRij{\ifmmode{[\Delta R_{ij}]}\else\hbox{$[\Delta R_{ij}]$}\fi}

\def\ns{\ifmmode{N_{\rm s}}          
        \else\hbox{$N_{\rm s}$}\fi}

\def\mat#1{{\bf #1}}     
\def\vek#1{{#1}}         

\newcount\eqcount
\def
  \nummer{
    \global\advance\eqcount by 1
    (\the\eqcount)
  }

\def
  \numadv{
    \global\advance\eqcount by 1
  }

\def
   \numout#1{
     (\the\eqcount #1)
  }

\def\ivek#1#2{\ifmmode{\vek{I}^{#1}_{#2}}
        \else\hbox{$\vek{I}^{#1}_{#2}$}\fi}

\def\tmat#1#2{\ifmmode{\mat{t}^{#1}_{#2}}
        \else\hbox{$\mat{t}^{#1}_{#2}$}\fi}
\def\rmat#1#2{\ifmmode{\mat{r}^{#1}_{#2}}
        \else\hbox{$\mat{r}^{#1}_{#2}$}\fi}
\def\bvek#1#2{\ifmmode{\beta^{#1}_{#2}}
        \else\hbox{$\beta^{#1}_{#2}$}\fi}

\def\bpi#1{\bvek{+}{#1}}

\def\lp{\ifmmode{\lambda^+_\tau}           
        \else\hbox{$\lambda^+_\tau$}\fi}
\def\lm{\ifmmode\lambda^-_\tau             
        \else\hbox{$\lambda^-_\tau$}\fi}

\newcommand{\fep}{Fe~II}

\begin{document}
 
\bibliographystyle{natbib-apj}

\citestyle{aa}
 
\title{NLTE Effects in Modeling of Supernovae near Maximum Light}

\author[E.~Baron et al.]{E.~Baron$^{1}$, P.~H.~Hauschildt$^{2}$,
P.~Nugent$^{1}$, and D.~Branch$^{1}$\\
$^{1}${Dept. of Physics and Astronomy, University of Oklahoma, 440 W.
Brooks, Rm 131, Norman, OK 73019-0225.}\\ $^{2}${Dept. of Physics and
Astronomy, Arizona State University, Tempe, AZ 85287-1504.}}

\maketitle
 
\begin{abstract} 
Supernovae, with their diversity of compositions, velocities, envelope
masses, and interactions are good testing grounds for probing the
importance of NLTE in expanding atmospheres. In addition to treating
H, He, Li~I, O~I, Ne~I, Na~I, and Mg~II in NLTE, we use a very large
model atom of Fe~II to test the importance of NLTE processes in both
SNe Ia and II. Since the total number of potential line transitions
that one has to include is enormous ($\approx 40$ million),
approximations and simplifications are required to treat the problem
accurately and in finite computer time. With our large Fe~II model
atom (617 levels, 13,675 primary NLTE line transitions) we are able to
test several assumptions for treating the background opacity that are
needed to obtain correct UV line blanketing which determines the shape
of near-maximum light supernova spectra. We find that, due to
interactions within the multiplets, treating the background lines
as pure scattering (thermalization parameter $\epsilon = 0$) is a poor
approximation and that an overall mean value of $\epsilon \sim 0.05 -
0.10 $ is a far better approximation. This is true even in SNe Ia,
where the continuum absorption optical depth at 5000~\AA\ ($\equiv
\tau_{\rm std}$) is $<< 1$. We also demonstrate that a detailed
treatment of NLTE effects is required to properly determine the
ionization states of both abundant and trace elements.
\end{abstract}

\section{Introduction}

Modeling  the light curve and spectrum of a supernova is
in principle a daunting task, since it requires full NLTE radiation
hydrodynamics in three spatial dimensions and detailed understanding
of the explosion mechanism. The dynamic range and computer power for
such detailed modeling does not currently exist. In practice one can
do spherically symmetric, LTE, multi-group radiation hydrodynamics
\cite[e.g.][]{hofsn94d} to determine the light curve, or one can calculate
steady-state detailed model atmospheres
\cite[e.g.][]{b93j3,nug1a95}. In fact, until details of the
explosion mechanisms are worked out and one gains the ability to
account for complicated micro and macroscopic mixing, these
approaches are both reasonable and accurate enough to extend our
understanding of the nature and systematics of supernovae, although in
principle one would like to relax the LTE assumption in the 
light curve models. Even within this somewhat limited framework, treating
all 40 million radiative transitions that occur in detailed NLTE is beyond the
capability of presently available computers,
and there is always the caveat that the atomic data is likely to be
inaccurate for many transitions. Nevertheless, through the use of a
non-local approximate lambda operator we have been able to include (on
workstation class computers) very detailed model atoms of iron group
elements \cite[]{phhs392,phhcas93,hbfe295}. Our aim in this paper is to examine the
effects of using such model atoms on the synthetic spectra of both Type
Ia and Type II supernovae and to use the knowledge gained from  
numerical experiments to guide our necessarily approximate treatment
of the the additional million or so radiative rates that are required
to correctly reproduce the UV line blanketing that occurs in
supernovae near maximum light. Actually, the task is even more
difficult, while including about 500,000 lines saturates the opacity
and the overall spectral shape and detailed lineshapes are well
reproduced, {\em which\/} 500,000 transitions are the most important
varies with supernova and phase and therefore must be dynamically
determined from the full list of 40 million possible transitions.


  We  briefly describe the \fep\ model atom
and the basic features of supernova model atmospheres \cite[details
can be found in][]{b93j3,nug1a95,hbfe295}. Some of the description of
the methods in Sections 2 and 3 has appeared in a companion paper on novae,
\cite[]{phhnovfe296} although the details are presented in those sections for
supernova models as opposed to nova models. The reader familiar with
\cite{phhnovfe296} may wish to examine the figures in those sections
and then skip to Section 4 where we  discuss results we
have obtained for representative supernova model atmospheres. We
conclude with a summary and discussion.

\section{Methods and Models}

\subsection{Radiative transfer}

In the Lagrangian frame (``comoving frame''),
the special relativistic equation of radiative transfer is given by
\cite[eg.,][]{found84}
\begin{eqnarray}
%
  \g (\mu + \b) \pder{I}{r}
%
   &  + \pder{}{\mu}\left\{ \g (1-\mu^2)
      \left[ {(1+\b\mu) \over r}
                - \g^2(\mu+\b) \pder{\b}{r} \right] I \right\} \nonumber \\
  &
    - \pder{}{\n} \left\{ \g
       \left[ \div{\b(1-\mu^2)}{r} + \g^2\mu(\mu+\b)\pder{\b}{r} \right]
         \n I \right\} \nonumber \\
  &
    + \g\left\{
       \div{2\mu+\b(3-\mu^2)}{r} + \g^2(1+\mu^2+2\b\mu)\pder{\b}{r}\right\}
      I \nonumber \\
  & \qquad =
      \tilde\eta - \chi I.\label{RTE}
\end{eqnarray}
Here, $r$ is the radius, $\m$ the cosine of the angle between a
ray and the direction normal to the surface, $\n$ the frequency,
$I=I(r,\m,\n)$ denotes the specific intensity
at radius $r$ and frequency $\n$ in the direction $\arccos(\m)$ in the
Lagrangian frame.
The matter velocity $v(r)$ is measured in units of the speed of light
$c$: $\b(r)=v(r)/c$ and $\g$ is given by $\g(r)=1/\sqrt{1-\b^2}$.
The sources of radiation present in the
matter are described by the emission coefficient
$\tilde\eta=\tilde\eta(r,\n)$,
and $\chi=\chi(r,\n)$ is the extinction coefficient.

We solve Eq.~(\ref{RTE}) using an operator splitting method with an
exact non-local 
band-matrix approximate $\Lambda$-operator \cite[]{phhs392,hsb94}.
 We can safely neglect
the {\em explicit} time dependencies and {\em partial} time derivatives
$\partial/\partial t$ in the radiative  transfer because the
radiation-flow timescale is much smaller than the timescale of the
evolution of 
the supernova outburst (at least near maximum light). However, the advection
and aberration terms must be retained in order to obtain a physically
consistent solution \cite[]{found84,per87,per91a,per91b}. This approach is also consistent
with the equation of radiation hydrodynamics in the time independent
limit \cite[]{found84,bhm96}.  We neither neglect the Lagrangian time
derivative $D/Dt = \partial/\partial t + v \partial/\partial r$,  nor
assign an ad hoc value to $D/Dt$  \cite[]{eastpin93}.
The latter two assumptions lead to physical
inconsistencies with the equations of radiation hydrodynamics (e.g.,
they do not lead to the correct equations for a steady-state stellar
wind). Our approach is physically self-consistent because it includes
the important advection and aberration terms in both the non-grey
radiative transfer and the radiative energy equations. The {\em only}
term that we neglect is the {\em explicit} time dependence, which is a
very good approximation in nova and supernova atmospheres \cite[]{bhm96}.

Although our method is much more complicated than ad hoc assumptions for 
$D/Dt$ (because of the additional terms in the equations that must be handled 
that break the symmetry of the characteristics of the radiative transfer 
equation), its results are much more reliable than those of simpler methods. 
In addition, the solution of the correct set of radiative transfer and energy 
equations in the co-moving frame is no more time consuming than the solution 
of the corresponding static problem. This is because of our use of a non-local 
approximate $\Lambda$-operator. 

\subsection{Treatment of spectral lines}

The physical conditions prevailing in nova and supernova atmospheres
are such that a large number of spectral lines are present in the line
forming regions. Therefore, the simple Sobolev approximation {\em
cannot} be used for accurate modeling including both line and continuum
transfer, because many lines of comparable strength overlap
\cite[see][]{phhnov95}. This means that the basic assumptions required
to derive the Sobolev approximation are not valid. We demonstrate this
in Fig.~\ref{linefig} in which we plot the number of lines that are
stronger than the local b-f continuum and that lie within a $\pm 2$
Doppler-width wavelength interval around each wavelength point in the
co-moving frame. This graph is for a representative Type Ia supernova
atmosphere [W7 abundances \cite[]{nomw7}, homogenized for $v>
8,000$~\kms, see e.g. \cite{nug1a95}] with an effective temperature of
$9,000\,$K, and a micro-turbulent or statistical velocity of
$\xi=2\kms$. In the UV the number of overlapping strong lines at each
wavelength point is typically larger than 100, in some regions as many
as 300 strong lines lie within 2 Doppler-widths. Even in some regions
of the optical spectral range, the number of overlapping lines can be
as high as 200 or more. Fig.~\ref{linefig} shows that the situation
becomes much worse for weaker lines (lines that are stronger than
$10^{-2}$ of the local b-f continuum must be included in supernova
atmosphere models).  Now, nearly 1000 lines (all of comparable
strength) overlap in the UV and around 300 lines in the optical can
overlap at each wavelength point. This shows decisively that the
Sobolev approximation cannot be used in modeling supernova
atmospheres. In light of this observation,  our approach  solves the
full radiative 
transfer equation for all lines. In addition, we wish to model stellar
atmospheres (both static cool stars, and hot stars with winds) and in
those cases the Sobolev approximation does not apply.

\begin{figure*}
\caption{\label{linefig}The lower line shows the number of overlapping
lines within two Doppler widths of each wavelength point when lines
are included that are at least as strong as the bound-free
continuum. The upper line shows the same when weak lines that are
within 0.01 of the bound-free continuum are included.}
\end{figure*}

\section{The Fe~II model atom}

We have taken the data needed to construct our model atom from the
compilation of
\cite{cdrom22}, whose long term project to provide 
accurate atomic data for modeling stellar atmospheres is an invaluable service 
to the scientific community. For our current model atom, we have kept terms up 
to the $^2${H} term, which corresponds to the first 29 terms of \fep. 
Within these terms we include all observed levels that have observed b-b 
transitions with $\log{gf} > -3.0$ as NLTE levels (where $g$ is the 
statistical weight of the lower level and $f$ is the oscillator strength of 
the transition). This leads to a model atom with 617 levels and 13675 
``primary'' transitions which we treat in detailed NLTE. We solve the complete 
b-f and b-b radiative transfer and rate equations for all these levels and 
include all radiative rates of the primary lines. In addition, we treat the 
opacity and emissivity for the remaining nearly 1.2 million ``secondary'' b-b 
transitions in NLTE, if one level of a secondary transition is included in our 
detailed model. 

 Using this procedure to select our model atom, we obtain 13675
primary \fep\ NLTE lines between the 617 \fep\ levels included in
NLTE. For \fep\ 
lines with $\lambda > 3500\ang$, we use 5 to 11 wavelength points
within their profiles. In extensive test calculations, we have found
that \fep\ lines with $\lambda < 3500\ang$ do not require these
additional wavelength points in their profiles due to the fine
wavelength grid used in the model calculations at these wavelengths
\cite[]{hbfe295}. This procedure typically leads to about 30,000
wavelength points for the model iteration and the synthetic spectrum
calculations. Figure~\ref{sample} illustrates the adequacy of this
procedure. We compare the output spectra of two models, which were
iterated independently. One has the line by line opacity sampling
below $3500\ang$ and the other has full line profiles for all NLTE
lines. Clearly this wavelength grid is sufficiently dense for the
calculations.  This is mostly due to the properties of the \fep\ ion,
in particular that its lines are concentrated mostly in the wavelength
range below $3500\ang$. Other model atoms, e.g., Ti~I, require
additional wavelength points for practically every line (Hauschildt
\etal, in preparation). For all primary lines the radiative rates and
the ``approximate rate operators'' \cite[see][]{phhcas93} are computed
and included in the iteration process.
A detailed description of the \fep\ line treatment 
is presented in \cite{phhnovfe296}.

\begin{figure*}
\caption{\label{sample}The solid line shows the result of a
calculation of an SN~Ia model with every line profile function is
included with 5 points per line for a total of 83269 wavelength points
and the dashed line shows the results of sampling for $\lambda <
3500$~\ang, with a total of 38465 wavelength points.}
\end{figure*}

\section{Results}

In this section we compare and contrast the effects of NLTE \fep\ in
SNe~II, where iron is only a trace element with those in SNe~Ia where
iron is a major constituent. We will discuss the transition class
SNe~Ib/c in future work. Table~\ref{tab1} describes the basic model
parameters for the spectra that we use to illustrate the NLTE effects.

\newcommand{\dayonethree}{SN~87A\_day\_13}
\newcommand{\daythreeone}{SN~87A\_day\_31}
\newcommand{\sniamod}{SN~Ia}
\begin{table*}
\caption{Parameters of  Presented Models\label{tab1}}
\medskip
\begin{tabular}{lrlrrlcc}
Model &\Tmod&\Rzero&\vno&\xni&Abund&$N$&$v_e$\\
\dayonethree&5400&$6.1\times 10^{14}$~cm&5500~\kms&0.03&$0.33\,Z_\odot$&6&---\\
\dayonethree&5100&$4.5\times 10^{14}$~cm&1700~\kms&0.01&$0.4\,Z_\odot$&6&---\\
\sniamod&9000&$9.7\times 10^{14}$~cm&7500~\kms&0.05&W7&---&900~\kms\\
\end{tabular}

\medskip
The basic model parameters for the spectra presented. The model
atmospheres are characterized by the following parameters
\cite[see][for details]{hwss92,b93j3}:   (ii) the
model temperature \Tmod\unskip, which is defined by means of the
luminosity, $L$, and the reference radius, $\Rzero$, ($\Tmod=(L / 4\pi
\Rzero^2 \sigma )^{1/4}$ where $\sigma$ is Stefan's constant); the
reference radius 
$\Rzero$, which is the radius where the continuum optical depth in
extinction at 5000~$\ang$ ($\tau_{\rm std}$) is unity;
the
expansion velocity, $\vno$, at the reference radius; the mass fraction
of \nni, \xni, which is the source of non-thermal gamma-rays; the
elemental abundances
which are for the SN~1987A models solar \cite[]{ag89} with metalicity
scaled by the factor given in the table and the SN~Ia model
are  W7
\protect\cite[]{nomw7} compositions, homogenized for $v>8,000$~\kms.
the density structure parameter, $v_e$, ($\rho(r)\propto
\exp(-v/v_e)$), or $N$ ($\rho(r)\propto (v/\vno)^{-N}$)
\end{table*}

\subsection{SNe II}

In SNe~II, iron is only a trace element in the outer ejecta, and thus,
one might expect that its effects on spectrum formation would be
small. This is not the case, since iron line blanketing still plays
the dominant role in the formation of the
spectrum. Figures~\ref{day13_comp}~and~\ref{day31_comp} display
calculations with Fe~II treated in LTE and NLTE compared to observed
spectra of SN~1987A 13 and 31 days after explosion. The
varying role of \fep\ NLTE is nicely displayed in these figures. While
we have attempted to fit the observed spectrum, neither the LTE nor
NLTE models match the width of the all of the observed features, since
composition inhomogeneities as well as more complicated density
structures (including clumping) are known to be important in
SN~1987A. Nevertheless, the NLTE \fep\ models overall, particularly in
the blue, provide a
better fit to the observed spectra than do the LTE \fep\ models.

In Fig~\ref{day13_comp} the H$\alpha$ and H$\beta$  absorption is too
narrow in both 
models and the absorption feature at $5800\ang$, which is likely due
to He~I $\lambda 5876$ and the Na D doublet, is poorly fit. However,
below $\lambda \sim 4200\ang$, the NLTE model more closely follows the
observed spectrum. In the later spectrum (Fig.~\ref{day31_comp}), the
overall fit of the NLTE model is again superior to that of the LTE
spectrum, even though the Balmer lines are now quite complicated
probably due to composition inhomogeneities. 

\begin{figure*}
\caption{\label{day13_comp}The observed spectrum of SN 1987A (taken at CTIO) on
March 8, 1987 (thin solid line) is compared to two model calculations
with reference radii and velocities of $6.1 \times 10^{14}$~cm and
$5500$~\kms\ and $\Tmod=5400$~K, $\xni=0.03$, and $Z=0.33\,Z_\odot$. The
density profile is a powerlaw with $N=6$ (model \dayonethree\ in
Table~\protect\ref{tab1}). The 
thick solid line includes Fe~II in NLTE, whereas the thin dot-dashed
line displays a calculation where Fe~II is treated in LTE.}
\end{figure*}

\begin{figure*}
\caption{\label{day31_comp}The observed spectrum of SN 1987A (taken at CTIO) on
March 26, 1987 (thin solid line) is compared to two model calculations
with reference radii and velocities of $4.5 \times 10^{14}$~cm and
$1700$~\kms\ and $\Tmod=5100$~K, $\xni=0.01$, and $Z=0.4\,Z_\odot$. The
density profile is a powerlaw with $N=6$ (model \daythreeone\ in
Table~\protect\ref{tab1}). The
thick solid line includes Fe~II in NLTE, whereas the thin dot-dashed
line displays a calculation where Fe~II is treated in LTE.}
\end{figure*}

\subsection{SNe~Ia}

In SNe~Ia, where iron is a major constituent of the atmosphere, one
expects that \fep\ NLTE effects would play a dominant
role. Figure~\ref{sn1a_nlte_lte}a compares the results of two
representative SNe~Ia model calculations where \fep\ is treated in LTE
and NLTE. In both cases the models have been iterated to radiative
equilibrium. In order to better display the differences between the
two calculations, Fig.~\ref{sn1a_nlte_lte}b displays the
differences between the two spectra, where a positive value implies
more flux in the NLTE model.  The differences are significant,
particularly since one goal of  spectral analysis through synthetic
spectra is to determine 
abundances in supernovae.

Up to this point when we have referred to LTE treatment, we have
really meant transitions, for which the rates are not included in the
solution of the rate equations. However, the source function is {\em
not\/} treated in complete LTE in that we write the source function in the
line as:
\[
S_\lambda = \epsilon\,B_\lambda + (1-\epsilon)\,J_\lambda,
\]
where $\epsilon$ is the thermalization parameter (which we take to be
a constant in the range $0.05\le\epsilon\le0.1$) and the other symbols
have their usual meaning. In a pure LTE treatment $\epsilon=1$, and in
a ``pure scattering'' treatment $\epsilon=0$. Figure~\ref{scat}
compares the results of a full NLTE treatment, a pure LTE treatment,
and a pure scattering treatment of the lines, with the structure held
fixed. The results are quite interesting. The pure LTE case reproduces
the overall spectral shape of the NLTE case rather well, while the
pure scattering case does not. In fact the overall luminosity differs
in the NLTE case and the pure scattering case by more than a factor of
2. The lineshapes, are however, better reproduced in the pure
scattering case.  This result is not surprising, since in many lines
an analytic estimate of $\epsilon$ such as given by the equivalent two
level atom formulation would predict a very small 
value of $\epsilon \la 0.01$. On the other hand, the total shape of
the spectrum depends on the interaction between lines and continua as
well as the interaction among overlapping lines (i.e. blanketing, blending, and
collisions in the multiplets). The dotted line shows a case where only
594 b-b transitions have been included and illustrates that the ``pure
scattering'' approximation does not correctly pass flux from blue to
red. Thus, even though it reproduces the lineshapes in the optical,
and since it has the opacity to correctly reduce the flux due to line
blending, the flux is simply scattered and not passed to longer
wavelengths as it should be. In the NLTE case, the multitude of
collisions within multiplets \cite[see also][]{hofasi96} creates a
pseudo-thermal 
pool of photons which are in equilibrium among themselves and have a
more nearly isotropic momentum distribution.

\subsection{NLTE effects on the Fe~II ionization}

In Fig.~\ref{bground} we display the ground state departure
coefficients, $b_1$, for two supernova models. Since supernovae
produce significant amounts of radioactive \nni\ it is important to
include the effects of non-thermal electron (produced by $\gamma$-ray
deposition) collisions in the NLTE rate equations. In the models
presented here the $\gamma$-ray deposition is treated as local due to
a nickel mass fraction \xni,  described in Table~\ref{tab1}.
The effects of non-thermal collisional ionization by primary electrons
produced by collisions with gamma rays due to the decay of $^{56}$Ni
and $^{56}$Co are modeled using the continuous slowing down
approximation \cite[]{gg76,swartz91,b94i1}. We have also included
ionizations due to secondary electrons \cite[]{swartz91,b94i1}, but most of
their energy is thermalized and therefore has only a small effect on
the level populations \cite[]{meyerott80}. The collisional cross
sections are taken from the work of 
\cite{lotz67a,lotz67b,lotz68a,lotz68b,lotz68c}.

Figures~\ref{bground}a,b show the results for the model \sniamod\ 
 with and without non-thermal effects,
whereas Fig.~\ref{bground}c,d shows the results for the model
\dayonethree, with 
and without non-thermal effects.  We have marked 
the $b_1$ for Fe~II with connected plus-signs for clarity. In all
models, the departures from LTE are significant.  In most regions of
the supernova atmospheres, the Fe~II $b_1$ are smaller than
unity. This usually indicates an overionization of Fe~II relative to
LTE, but we will see below that this is not always the case.  The
departure coefficients of other species show the same behavior as
Fe~II, which is very different from the results we have obtained in
our nova atmosphere modeling \cite[]{phhnovfe296}. The sharp drop of
the departure coefficients of He~I in the model \dayonethree\ 
is caused by the ionization of helium by non-thermal electrons in this
model as can be seen by comparing the models with and without
non-thermal effects. Most effected by the non-thermal effects are
He~I, \fep,\ and Mg~II. Ca~II appears to be hardly affected. Comparing
the SN~Ia models with and without non-thermal effects, it is evident
that the non-thermal effects play an important role in determining the
level populations. Again Ca~II, appears to be the least affected by
non-thermal effects.  In the SN~II model O~I is not strongly affected
by non-thermal collisions although that is not the case in the SN~Ia
model.  In the SN~Ia model, NLTE effects are quite small when
non-thermal ionization is neglected. This may be due to the fact that
since SN~Ia atmospheres consist entirely of metals, with lower
ionization energies than hydrogen and helium, that thermal electron
collisions act to restore LTE. Once all important non-thermal electron
collisons are included NLTE effects become important.

If Fig.~\ref{fe-ion-snii} we display the ionization structure of iron
throughout the atmosphere for model \dayonethree. In panels (a and b) we show the
NLTE results (with and without non-thermal electrons) whereas panels
(c and d) show the results for LTE (but otherwise the same model
structure). First note the number of iron ionization stages that are
present in the atmospheres (in a number of interleaved ``Str\"omgren
Spheres''). This is a more pronounced and well-known feature of nova
atmospheres which is caused by the large temperature gradients inside
the expanding shell. We find similar effects for all elements that are
included in the model calculations.

 The effects of NLTE on the Fe~II ionization balance become clear by
comparing panels (a and c) and (b and d) in Fig.~\ref{fe-ion-snii}. In
the LTE case, 
Fe~II is the dominant ionization stage of iron in the outer atmosphere,
In the NLTE-Fe~II case, however, Fe~II recombines into Fe~I throughout
the outer atmosphere, although the ground state departure coefficients
of Fe~II are less than unity (the latter fact usually indicates a
overionization relative to LTE). This is true whether or not
non-thermal effects are included, although the iron remains ionized to
lower optical depths when non-thermal effects are included. The reason
for this non-intuitive ionization 
behavior becomes clear by inspecting the graphs shown in
Fig.~\ref{ion-struc-snii}. In the NLTE case (Fig.~\ref{ion-struc-snii}a),
hydrogen remains ionized throughout the outer atmosphere, and the electron
densities therefore remain high. However, in LTE, hydrogen recombines and
the electron densities are much smaller than in the NLTE model. This 
completely changes the behavior of the ionization equilibrium for {\em all}
species and the higher electron pressures combined with the different
recombination rates in the NLTE model force the 
recombination of Fe~II to neutral iron. This effect illustrates that 
NLTE modeling is both extremely important and must be done in great
detail in order to obtain realistic results. In particular, this shows
that to fit an observed  SN~II using a model in LTE,  would force the
choice of a high 
temperature in order to reproduce the Balmer lines. With this high
temperature, in LTE the ionization of Fe could never be reproduced.
The correlation of the  recombination of iron  with the ionization of
hydrogen is evident upon comparing 
Fig.~\ref{ion-struc-snii}a and Fig.~\ref{ion-struc-snii}c. In panel
(a) which includes non-thermal electron collisions, the iron is
neutral deeper into the atmosphere, whereas in panel (c) without
non-thermal effects the iron is only recombining in the outer regions
just where the electron and H~II fractions are rising.

In Figure~\ref{fe-ion-sni} we show that the changes in the iron
ionization structure in the SN I model are very small. This is
expected since the ground state departure coefficients for Fe~II are
close to unity in this model. Again, with the neglect of non-thermal
effects, the high electron fractions in SN~Ia models drive them
towards LTE. This is very different from the behavior we find in both
the SN II model and our nova model atmospheres \cite[]{phhnovfe296}.
The NLTE effects of Fe~II in the SN I model will be apparent only in
the line source functions and the spectra, as discussed below.

 In Fig.~\ref{allbi} we show the departure coefficients for all \fep\
levels for the two models (with and without non-thermal effects) in an
overview graph. This figure demonstrates that the NLTE effects for
\fep\ are important and differ for the individual levels and for
different model  temperatures, \cite[for a direct comparison of
effect of varying the model temperature in SNe~II,
see][]{hbfe295}.  For each $\Tmod$, the departure coefficients are
closer to unity (their LTE value) for the higher lying levels, which
are more strongly coupled to the continuum than the lower lying
levels.  The rather large variation of the $b_i$ within the \fep\
model atom suggests that the \fep\ lines will show significant NLTE
effects themselves, in addition to the effects introduced by the
changes in the ionization structure.  Comparing the models with and
without non-thermal effects we see that the models without non-thermal
ionization have departure coefficients larger than unity at moderate
optical depths, however when non-thermal ionization is considered the
departure coefficients tend to stay below one.

\subsection{NLTE effects on the formation of Fe~II lines}

The NLTE effects not only change the ionization balance of iron but
also change the relative level populations of Fe~II. We demonstrate
this effect for two important \fep\ multiplets. In Figs.~\ref{uv1}a,b
we show the ratio of the line source functions $S_L$ to the local
Planck function for the UV1 multiplet of \fep. These are UV lines
(around $2600\ang$) from the ground $\rm a ^6D$ term to the $\rm z
^6D^o$ term. In the model \sniamod\ which includes non-thermal effects
(Fig.~\ref{uv1}a, left panel) the UV1 multiplet is nearly in LTE up to
the outer atmosphere. Only below $\tstd=10^{-3}$ do the line source
functions differ, by about 20\%, from the Planck function.  The
situation is different for the model \dayonethree\ (Fig.~\ref{uv1}a, right
panel).  Here, the line source functions are larger than the Planck
function by nearly a factor of 2. In both models, the (collisional)
coupling between the levels of each term is strong enough to nearly
establish the same line source function to Planck function, $S_L/B$,
ratio for all lines within the multiplet, although the electron
densities are relatively low. Only in the SN II model do the levels
within a term finally decouple for $\tstd < 10^{-4}$.  When
non-thermal processes are neglected, the same general trend is
observed although now the SN~II model has a UV1 line source function
that is smaller than the local Planck function inside, but becomes
greater than the local Planck function far out in the atmosphere.

 The subordinate \fep-multiplet 42 ($\rm a ^6S$ to $\rm z ^6P^o$)
shows a somewhat different behavior in Fig.~\ref{uv1}a. The line
source functions for the 3 primary NLTE lines are slightly smaller
than unity for $10^{-4} \le \tstd < 2$ and larger than unity for
$\tstd < 10^{-4}$ for the model \sniamod\ (left panel). The model
\dayonethree\ (right panel) shows nearly the opposite behavior. The NLTE
effects are slightly larger for the cooler model, as seen earlier for
the UV1 multiplet. This shows that the NLTE effects for \fep\ are not
just over-ionization but that the ``internal'' NLTE effects are also
significant. This multiplet is less affected by non-thermal ionization
than is the UV1 multiplet. 

 The Fe~II NLTE effects we have found in SN I and II atmospheres are
significantly {\em smaller} than in nova atmospheres \cite[]{phhnovfe296}.
This is due to the much higher velocity field in
SN atmospheres, which tends to transfer photons from the regions
neighboring the Fe~II lines into them, thus coupling the frequencies
stronger together than in the case of novae. This is also true for the
regions around the ionization edges of Fe~II, which limits the effective
number of Fe~II ionizing photons. It must be noted here that our
departure coefficients, radiation field and source functions are {\em
all} computed in the co-moving frame and, therefore, simplified
arguments based on Eulerian frame experience can and will lead to wrong
conclusions. In the co-moving frame, the velocity does {\em not} lead to
a frequency shift of the photons (by its construction) but it leads
instead to terms which are better described as ``drag'' terms, slowing
the response of the radiation field to changes in the opacity as a
function of wavelength when  compared to the static case. This
``stiffer'' radiation field limits the magnitude of the NLTE effects on
the line formation, in fact this is more effective at higher velocities.
Therefore, stellar winds and novae show comparatively larger NLTE
effects than the very fast expanding atmospheres of SNe.

\section{Conclusions}

We have demonstrated that large model atoms are required in order to
correctly transport energy and push the flux from blue to red in SNe
atmospheres. Weak lines must be included in order to correctly
reproduce line blanketing and blending. In spite of the fact that an
equivalent 2-level atom approach would predict very small values of
the thermalization parameter $\epsilon$ for many transitions,
collisions in the multiplets and continuum effects redistribute the
energy and momentum such that an overall higher value of $\epsilon$
leads to a more accurate description of the radiation transport.
A calculation that treats weak lines as pure scattering misses far
too much opacity and gives results similar to those that treat only
continua. Accurate modeling of the UV also requires large model atoms
and the inclusion of weak lines as background opacity. As better
atomic data become available the accuracy of the calculations should
improve. 
NLTE effects cause very non-intuitive ionization effects to occur
which show that these effects must be included in detail.

\section*{Acknowledgements}
We thank the referee for suggestions which improved the presentation
of this paper.
This work was supported in part by NASA grants NAGW-4510, NAGW-2628,
and NAGW 5-3067 to Arizona State
University, NSF grants AST-9417242 and
AST-9115061; and an IBM SUR grant to the University of Oklahoma.  Some
of the calculations presented in this paper were performed at the
Cornell Theory Center (CTC), and the San Diego Supercomputer Center
(SDSC), supported by the NSF, and at the National Energy Research
Supercomputer Center (NERSC), supported by the
U.S. DoE, we thank them for a generous allocation of computer time.

\bibliography{refs,rte,sn1a,atomdata,crossrefs}

\begin{figure*}
\caption{\label{sn1a_nlte_lte}Panel (a):The thick solid line corresponds to a
SN~Ia model where \fep\ is treated in NLTE, and the thin solid line
corresponds to a model where \fep\ is treated in LTE. Both models are
iterated to radiative equilibrium and have the parameters: W7
\protect\cite[]{nomw7} compositions, homogenized for $v>8,000$~\kms,
reference radii and velocities of $9.7 \times 10^{14}$~cm and
$7500$~\kms\ and $\Tmod=9000$~K, $\xni=0.05$ (model \sniamod\ in
Table~\protect\ref{tab1}).  The density profile is a exponential in
velocity with scale $v_e=900$~\kms. The thick solid line includes
Fe~II in NLTE, whereas the thin dot-dashed line displays a calculation
where Fe~II is treated in LTE. Panel (b):The relative difference
between the spectra shown in Panel (a), where a
positive value corresponds to more flux in the NLTE model.  }
\end{figure*}

\begin{figure*}
\caption{\label{scat}The results of a model calculation where the
lines are treated in full NLTE (thick solid), in pure LTE (thin
solid), in pure scattering (short-dashed), and where only the 594 b-b
transitions of the model atoms Mg~II, Ca~II, O~I, Na~I, Ne~I, H~I,
He~I, and He~II are included (long-dashed line).}
\end{figure*}


\begin{figure*}
\caption[]{\label{bground}The run of the ground state departure
coefficients $b_1$ for the NLTE species as functions of $\tstd$ (the
b-f optical depth at $5000\ang$). Panels (a,b) show the results
obtained for the model \sniamod, with and without non-thermal
ionization, respectively and panels (c,d) the model \dayonethree, with and
without non-thermal ionization respectively.}
\end{figure*}


\begin{figure*}
\caption[]{\label{fe-ion-snii}Ionization structure of iron as function of
$\tstd$ (the b-f optical depth at $5000\ang$) for the model
\dayonethree. Panel (a) shows the results obtained with Fe~II treated in
NLTE, with non-thermal ionization, panel (b), NLTE without non-thermal
ionization, whereas panel (c) shows the results for LTE Fe~II, with
the structure of the non-thermally ionized model, and panel (d) shows
LTE Fe~II, with the structure of the model neglecting non-thermal
ionization. The NLTE under-ionization causes Fe~II to recombine to
Fe~I in the outer layers of the SN atmosphere. Note the number of
``Str\"omgren Spheres'' of various ionization stages of iron in the
inner parts of the SN atmosphere.}
\end{figure*}


\begin{figure*}
\caption[]{\label{ion-struc-snii}The concentrations of the
most abundance species as function of $\tstd$ (the b-f optical depth
at $5000\ang$) for the model \dayonethree.  Panel (a) shows the results
obtained with in NLTE, including non-thermal ionization, panel (b),
NLTE without non-thermal ionization, whereas panels (c, d) show the
results for LTE, respectively. The NLTE under-ionization causes Fe~II
to recombine to Fe~I in the outer layers of the SN atmosphere. Note
the number of ``Str\"omgren Spheres'' of various ionization stages of
iron in the inner parts of the SN atmosphere. Non-thermal processes
prevent this recombination from occuring deeper in the atmosphere.}
\end{figure*}


\begin{figure*}
\caption[]{\label{fe-ion-sni}Ionization structure of iron as function of
$\tstd$ (the b-f optical depth at $5000\ang$) for the model \sniamod.
 Panels (a, b) show the results
obtained with Fe~II treated in NLTE, with and without non-thermal
ionization, respectively, whereas panels (c, d) show the results for
LTE Fe~II, with and without non-thermal ionization respectively. In
this model, the changes of the ionization of iron are 
very small and visible only in the outermost parts of the atmosphere.
Note the number
of ``Str\"omgren Spheres'' of various ionization stages of iron in the inner
parts of the SN atmosphere.}
\end{figure*}


\begin{figure*}
\caption[]{\label{allbi} The departure coefficients for all NLTE
levels of \fep\ 
as functions of the standard optical depth $\tstd$ for two models:
\sniamod, with and without non-thermal effects (panels a
and b) and \dayonethree, with and without non-thermal effects
(panels c and d). 
The
departure coefficients are closer to unity for the higher lying levels.}
\end{figure*}

\begin{figure*}
\caption[]{\label{uv1}The ratio of the line source function to the
local Planck function for the UV1 multiplet ($\rm a ^6D$ to $\rm z
^6D^o$) and for multiplet 42 ($\rm a ^6S$ to $\rm z ^6P^o$) as
functions of the standard optical depth. The left side displays the
model \sniamod\ and the right side the model \dayonethree. Panel a includes
non-thermal ionization and panel b does not.}
\end{figure*}


\end{document}